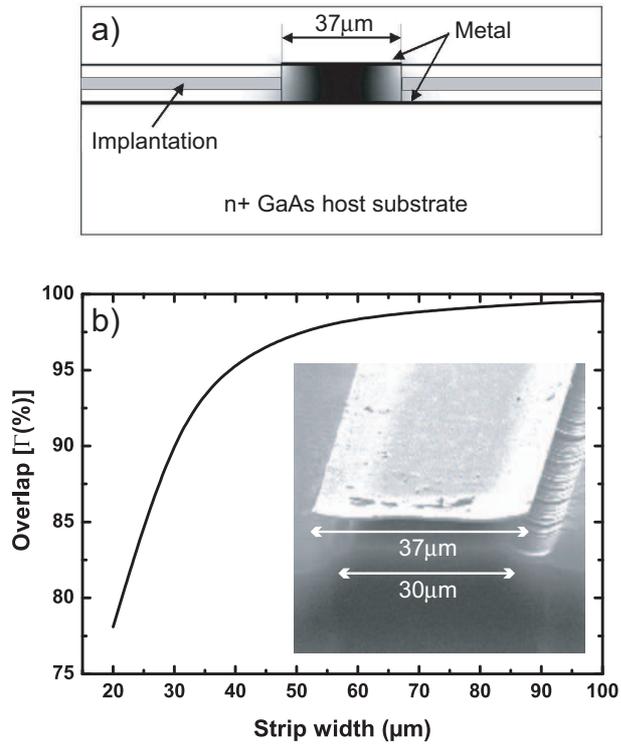

Figure 1. Dhillon *et al*

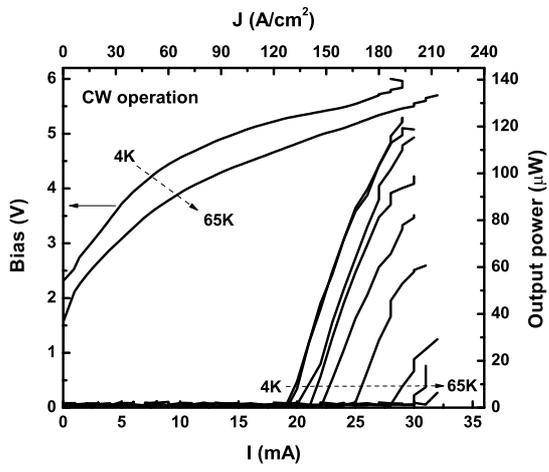

Figure 2. Dhillon *et al*

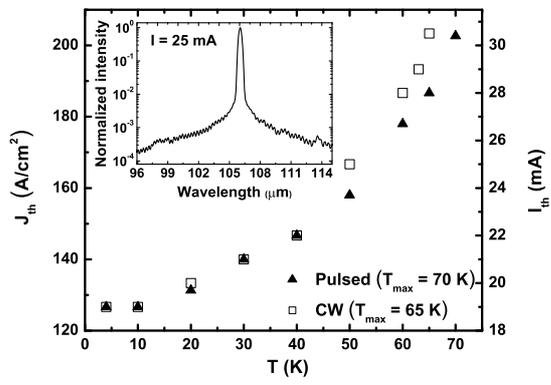

Figure 3. Dhillon *et al*

# Ultra low threshold current THz quantum cascade lasers based on buried strip-waveguides

Sukhdeep Dhillon[a]

Thales Research and Technology, 91404 Orsay, France

Matériaux et Phénomènes Quantiques, Université Denis Diderot - Paris 7, Paris, France

Jesse Alton[b], Stefano Barbieri

Teraview Ltd., Platinum Building, St John's Innovation Park, Cambridge, CB4 0WS, United Kingdom

Cavendish Laboratory, University of Cambridge, Madingley Road, CB3 0HE, United Kingdom

Carlo Sirtori

Thales Research and Technology, 91404 Orsay, France

Matériaux et Phénomènes Quantiques, Université Denis Diderot - Paris 7, Paris, France

A. de Rossi, M. Calligaro

Thales Research and Technology, 91404 Orsay, France

Harvey E. Beere, David Ritchie

Cavendish Laboratory, University of Cambridge, Madingley Road, CB3 0HE, United Kingdom

Abstract

THz quantum cascade lasers based on a novel buried cavity geometry are demonstrated by combining double-metal waveguides with proton implantation. Devices are realised with emission at 2.8 THz, displaying ultra low threshold currents of 19 mA at 4K in both pulsed and continuous wave operation. Thanks to the semiconductor material on both sides of the active region and to the narrow width of the top metal strip, the thermal properties of these devices have been greatly improved. A decrease of the thermal resistance by over a factor of two compared to standard ridge double-metal lasers of similar size has been measured.

---

[a] sukhdeep.dhillon@thalesgroup.com

Quantum cascade lasers (QCLs) represent a promising technology towards the realisation of room-temperature semiconductor coherent sources emitting at frequencies above 1 THz[1-5]. Although room-temperature operation is certainly desirable, in many cases the use of cryogenic equipment does not constitute a major obstacle. For example, lightweight, portable closed cycle Stirling coolers are routinely employed to operate quantum well infrared photodetectors at liquid nitrogen temperatures or below[6]. Typical cooling powers for this type of coolers are of a few 100 mW at the lowest temperatures, which can represent a severe constraint for QCLs, traditionally characterised by rather high power consumption. In this respect, compared to devices emitting in the near and mid-infrared, QCLs operating in the THz region of the electromagnetic spectrum are an exception, owing to their much lower threshold current densities. Yet, so far, owing to the large size of the devices imposed by the wavelength, or to high threshold voltages, the operating powers reported in the literature are always in excess of ~ 2 W[1-5].

In this work we employ a recently demonstrated bound-to-continuum QCL emitting at 2.8 THz, characterised by a threshold current density of only 110 A/cm$^2$ at 4K, together with a threshold voltage of just over 2V[4]. We take full advantage of these properties by incorporating the active region into a buried waveguide scheme based on double-metal technology[7,8]. Relying on proton implantation for current confinement, this technology allows the fabrication of devices of extremely small size, of the order of the wavelength in the material, with, in addition, considerably improved thermal handling capabilities compared to lasers based on standard ridge-cavities[7]. With a 500 x 37 µm device we demonstrate a threshold current of 19 mA in continuous wave at 4K, yielding a dissipated power at threshold of ~ 100 mW, which represents a performance record for ridge cavity QCLs.

We have recently demonstrated THz QCL buried waveguides based on surface plasmons[7]. We showed how a surface plasmon mode of finite vertical and lateral dimensions can be bound at the interface between a metal strip and a semiconductor. For a given wavelength, the width of the strip controls the degree of spatial confinement, with the optical mode expanding progressively inside the semiconductor as the strip width is reduced. By exploiting this concept we implemented single-plasmon buried waveguides where a ~ 12 µm thick active region lies on top of a semi-insulating substrate. Here the metal strip forming the top contact metallisation sustains the plasmon mode, which, in turn, strongly leaks inside the substrate by penetrating through a 700 nm thick, $2\times10^{18}$ cm$^{-3}$ n-doped bottom contact GaAs layer. At $\lambda = 100$ µm, we found that strips thinner than ~ 120 µm could not be used without reducing the overlap between the optical mode and the electrically pumped active area, therefore leading to higher threshold current densities. Yet, the advantage of realising even thinner laser guides without compromising the overlap would be twofold, yielding (i) devices operating at lower currents, and (ii) characterised by a higher heat conductance. This last point stems from the fact that in QCLs the excess heat is more efficiently dissipated in the direction parallel than perpendicular to the hetero layers, thanks to the higher in-plane heat conductivity resulting from bulk-like phonon dispersions[7,9]. Therefore we expect the realisation of devices with a better aspect ratio between lateral and vertical areas (i.e. with thinner strip widths) to result in improved thermal handling capabilities.

So-called double-metal waveguides are particularly suited for the realisation of small surface area THz QCLs[3,8]. In their original scheme the active region is sandwiched between two metal layers and the lateral optical and electrical confinement is obtained by etching through the active region down to the bottom contact layer. In Fig. 1(a) we present our alternative approach based on a buried waveguide geometry. Mode intensity calculations, based on a two-dimensional finite element modelling, were performed at $\lambda = 100$ μm, with gold as the surface metal. The active region used is described in Ref. [4] and is sandwiched between 700 nm $2\times10^{18}$cm$^{-3}$ and 80 nm $5\times10^{18}$cm$^{-3}$ doped layers for the lower and upper contacts, respectively.

As shown by the 2-D mode intensity plot, sub-wavelength optical confinement is achieved despite the total absence of a ridge, with the optical mode confined underneath a top-contact metal strip only 37µm wide. This results from the double-plasmonic nature of the optical mode, bound at both top and bottom metal-semicondctor interfaces, allowing for a much tighter lateral confinement compared to buried single-plasmon waveguides[1,3]. To confine the electrical current, high resistivity regions are realised on both sides of the top contact metal strip via proton implantation[7]. The presence of these regions is crucial for device operation; in fact, their elimination resulted into systematic device failure due to lateral current spreading, even after removing the highly doped top contact layer.

In Fig. 1(b) the modal overlap is plotted as a function of metal strip width, to show its effect on the lateral confinement (the overlap factor is calculated assuming an electron channel and metal strip of equal widths). The overlap starts to decrease for strip widths lower than ~ 40 µm: this must be compared with the 120 µm found for single plasmon buried waveguides[7]. The calculated waveguide losses ($\alpha_w$) are approximately constant ($\alpha_w$ ~ 24cm$^{-1}$) for strip widths between 20 and 100 µm.

Buried strip-waveguide devices were fabricated based on the scheme of Fig. 1(a). First, the active region of Ref. [4] was grown by Molecular Beam Epitaxy on top of a 300 nm $Al_{0.5}Ga_{0.5}As$ etch stop layer. By using the techniques detailed in references 3 and 8, the active region was then bonded on top of an n$^+$ GaAs host substrate, and the original semi-insulating substrate removed by wet etching. Subsequently, a 37µm wide Ti/Au (20nm/150nm) metal strip was evaporated onto the exposed active region and an electrical channel 30µm in width was defined by proton implantation (see the Inset of Fig 1(b)). This was obtained using proton energies of 300keV and 800keV to realise, on either sides of the strip, two insulating layers at depths of approximately 3µm and 8µm from the wafer surface. Contrary to the buried single surface plasmon waveguide, with this scheme it is no longer necessary to define a ridge for electrical contact purposes as back contacting is made through the n$^+$ host substrate[7]. We also note that the top contacts were left un-annealed. This was to avoid weakening of the In-Au wafer bond between the active region and host substrate, which starts to deteriorate at temperatures above 200°C[3,10].

Devices were cleaved into laser bars, indium soldered to copper mounts, gold wire bonded and characterized in a continuous flow liquid helium cryostat. The emitted light was collected using f/0.9 optics and refocused on a broad area room-temperature thermoelectric detector. The detector was calibrated using an absolute THz power meter from Thomas

Keating Instruments. For pulsed operation, a train of 1µs pulses with a period of 100µs was used, modulated with a 5Hz square envelope to match the response of the detector.

In Fig. 2 the continuous wave (CW) electrical and optical characteristics of a 500µm long, buried strip-waveguide THz QCL are shown at different heat sink temperatures. The device operates up to 65K, with a threshold current of 19 mA at 4K, the lowest, to our knowledge, of any ridge-cavity QCL regardless of the emission wavelength. The corresponding threshold current density is 127A/cm$^2$. The initial voltage step in the V-I curve, as well as the overall higher operating voltage compared to standard ridge devices, is due to the Schottky nature of the non-annealed Ti/Au top contact[3,4].

The inset of Fig. 3 shows a typical CW spectrum, recorded at 4K with a Fourier transform interferometer operated in rapid-scan mode at a maximum spectral resolution of 0.25cm$^{-1}$. The emission wavelength is centred at ~ 106 µm (2.8 THz), close to that of the design, and the laser remains mono-mode over the entire current operating range, with a side mode suppression ratio of 30 dB. Indeed, for this 500 µm long device we compute a round-trip frequency of ~ 80 GHz, corresponding to a longitudinal mode spacing of ~ 0.3 meV. This represents as much as ~ 1/4 of the spontaneous emission linewidth as measured from electroluminescence spectra of this active region, preventing multimode emission. We note that the double-metal waveguide geometry allows for short cavity devices to be realised without sacrificing laser performance in terms of threshold current density, due to the inherently high reflectivity of the cavity mirrors (based on the model of Herzinger *et al,*, we compute reflectivities in excess of 90% depending on device size) [11]. The drawback, however, is the relatively low output power, of ~ 100 µW at 4K, compared to devices based on single plasmon waveguides, yielding power levels in the tens of mW range[4,7].

The threshold current density as a function of temperature, for the same 500 µm long device of Fig. 2 is plotted in Fig. 3 for pulsed and CW operation: maximum temperatures are of 70K and 65K respectively. By comparing the change in temperature between the pulsed and CW characteristics with the power dissipated at threshold we derive a total thermal resistance of 28K/W for the present device[12]. For a 1.5 mm long laser we obtain a total thermal resistance of 10.8 K/W, more than a factor of two lower compared to the 23.3 K/W obtained from a standard ridge-cavity double-metal sample of the same surface area (50 µm x 1.1 mm, not shown). The decrease in thermal resistance for the buried strip-waveguide is a result of the extra material on either side of the electrically pumped active region, which, moreover, benefits from the asymmetry in thermal resistance of QCLs[7,9]. This last fact is demonstrated by observing that for the 120µm wide single-plasmon devices of Ref. [7], we found a decrease in thermal resistance of only 30% from the ridge-cavity to buried waveguide geometry. As discussed at the beginning of this work this is due to different aspect ratios between the lateral surface, determined by the thickness of the active region, and the area of the device.

Compared to double-metal ridge-cavity samples, buried strip-waveguide devices present a smaller current dynamical range (defined as the current density span between threshold and power saturation). This is a consequence of a reduction in the maximum current density, from ~ 260 A/cm$^2$ at 4K, for a double-metal ridge-cavity laser, to the 195 A/cm$^2$ of device presented in Fig. 2. The reduction of the current range, related to device fabrication issues, is still under investigation and limits the operating temperatures of the buried devices. Double-metal ridge lasers, 1.5 mm long and 50 µm wide, operated up to 90 and 70K in pulsed and CW mode respectively (not shown). These devices also exhibited a slightly lower threshold current density (117 A/cm$^2$ at 4K), compared to the 127 A/cm$^2$ of Fig. 2 in agreement with the different mode overlaps, of 95% and 88% for the 50 µm wide etched ridge and the present buried cavity respectively.

In summary we have shown that by combining double-metal waveguiding with proton implantation it is possible to realise buried waveguides of extremely small width, comparable to the wavelength in the material, without decreasing the modal overlap, nor increasing the optical losses. These buried strip-waveguide QCLs yielded record low CW threshold currents of 19mA at T = 4.2K (with a power dissipated at threshold of ~ 100 mW) and more than a factor of two reduction in the thermal resistance, compared to similar sized ridge-cavity double-metal devices.

This work was partially funded by the European Community through the Framework VI Integrated Project "TERANOVA". We also acknowledge support from the EPSRC (UK) and the Royal Society.

**Figure captions**

**Figure 1:** a) Computed two-dimensional THz mode profile of the 37 µm wide buried double-metal waveguide used in the experiments. The electron channel is 37 µm wide and is defined by proton implantation. The calculated modal overlap and waveguide losses are 95% and 24cm$^{-1}$ respectively. b) Effect of metal strip width on the overlap of the optical mode with the active region. The overlap is calculated assuming an electrical channel equal in width to the metal strip. Inset. SEM picture of the laser facet after partial wet-etching. The etch speed is larger in the proton implanted regions, resulting into a 30 µm wide step visible on the top surface, in correspondence with the non-implanted electrical channel. The etch solution is responsible for the undercut underneath the 37 µm wide metal strip. The difference in the widths of the metal strip and the electrical channel was unintentional and resulted into an overlap factor of 88%, compared to the nominal 95%.

**Figure 2:** CW electrical and optical characteristics of a 500 µm long, 37 µm wide buried strip-waveguide from 4K to 65K.

**Figure 3:** Threshold current density as a function of heat-sink temperature in continuous wave and pulsed operation for the same buried strip-waveguide laser of Fig. 2. Inset: Continuous wave spectra recorded at I = 25 mA and T = 4.2 K, using an FTIR operated in rapid scan mode with a resolution of 0.25cm$^{-1}$. The side mode suppression ratio is of 30dB.